%% file: main.tex
\documentclass[acmsmall]{acmart}
\usepackage{graphicx}  
\usepackage{subfig}

\usepackage{booktabs} 

\usepackage[ruled]{algorithm2e} 

\acmDOI{0000001.0000001}

\begin{document}
\title{Combining Crowd and Machines for Multi-predicate Item Screening}

\author{Evgeny Krivosheev}
\affiliation{%
  \institution{University of Trento, Italy}
}
\email{evgeny.krivosheevlast@unitn.it}

\author{Fabio Casati}
\affiliation{%
  \institution{University of Trento, Italy and Tomsk Polytechnic University, Russia}
}
\email{fabio.casati@unitn.it}

\author{Marcos Baez}
\affiliation{%
  \institution{University of Trento, Italy and Tomsk Polytechnic University, Russia}
}
\email{marcos.baez@unitn.it}

\author{Boualem Benatallah}
\affiliation{%
  \institution{UNSW, Sydney, Australia}
}
\email{boualem@cse.unsw.edu.au}


\renewcommand\shortauthors{omitted et al.}

\begin{abstract}

This paper discusses how crowd and machine classifiers can be
efficiently combined to screen items that satisfy a set of predicates.
We show that this is a recurring problem in many domains, present
machine-human (hybrid) algorithms that screen items efficiently and
estimate the gain over human-only or machine-only screening in terms
of performance and cost.
We further show how, given a new classification problem and a set of
classifiers of unknown accuracy for the problem at hand, we can
identify how to manage the cost-accuracy trade off by progressively
determining if we should spend budget to obtain test data (to assess
the accuracy of the given classifiers), or to train an ensemble of
classifiers, or whether we should leverage the existing machine
classifiers with the crowd, and in this case how to efficiently
combine them based on their estimated characteristics to obtain the
classification. We demonstrate that the techniques we propose obtain
significant cost/accuracy improvements with respect to the leading
classification algorithms.
\end{abstract}

%
%


%
%

%
%

\maketitle
\input{paper_body}

\bibliographystyle{ACM-Reference-Format}

\bibliography{sample-bibliography} 
\end{document}

%% file: paper_body.tex
\section{Background and Motivation}

A frequently occurring classification problem consists in identifying items that pass a set of \textit{screening tests} (filters).
This is not only common in medical diagnosis but in many other fields as well, from database querying - where we filter tuples based on predicates \cite{optimalWidom2014}, to hotel search - where we filter places based on features of interest \cite{Lan_dynamicfilter_hcomp17}, to systematic literature reviews (SLR) - where we screen candidate papers based on a set of exclusion criteria to assess whether they are in scope for the review~\cite{Wallace2017crowdMLshort}.
The goal of this paper is to understand how, given a set of trained classifiers whose accuracy may or may not be known for the problem at hand (for a specific query predicate, hotel feature, or paper topic), we can combine machine learning (ML) and human (H) classifiers to create a hybrid classifier that screens items efficiently in terms of cost of querying the crowd, while ensuring an accuracy that is acceptable for the given problem.
We focus specifically on the common scenario of \textit{finite pool} problems, where the set of items to screen is limited and where therefore it may not be cost-effective to collect sufficient data to train accurate classifiers for each specific case.
To make the paper easier to read and the problem concrete, we will often take the example of SLRs mentioned above, which is rather challenging in that each SLR is different and each filtering predicate (called \textit{exclusion criterion} in that context) could be unique to each SLR (e.g., "exclude papers that do not study adults 85+ years old").


The area of crowd-only and of hybrid (ML+H) classification has received a lot of attention in the literature. 
Research in crowdsourcing has identified how to best classify items given crowd votes~\cite{LiuWang_truelabel,Dong2013,whitehill2009whose}, including  work that focuses specifically on the \textit{screening} problem discussed here ~\cite{optimalWidom2014,Wallace2017crowdMLshort,Mortensen2016crowd,Krivosheev_hcomp17,Lan_dynamicfilter_hcomp17}.
Interesting ways to combine ML and H classifiers have also been proposed, for example by building ML classifiers that can also include crowd votes as features, whose purpose is not only that of providing better classification but also of weighing the value (vs the cost) of obtaining additional crowd votes~\cite{Kamar_2012_combining}. 
Other hybrid approaches ask the crowd to extract interesting patterns to be then fed to algorithms for classification, as opposed to relying on ML to do this~\cite{flock_2015,Carlos_pattern}. 

The approach we propose leverages the information provided by each kind of classifier (machine and human) to improve the effectiveness of the other kind so that they can be stronger together.
The algorithm is based on a probabilistic model that adapts to the specific characteristics of each item, screening filter,  workers' accuracy, and ML classifier available to identify \textit{what} to ask next for each item (which filter to test) and \textit{if} we should stop polling the crowd for a given item, either because we reached a decision or because we realize we cannot do so cheaply and confidently (that is, the problem of classifying that item is too hard for the crowd-machine ensemble). 
The latter point is particularly important: there may be items, filters, or classification problems in general where it is questionable whether machines and/or the crowd can classify with acceptable accuracy. 
If the algorithm can identify this subset of items and filters - or (item,filter) pairs - early and avoid spending money on them, then we know we can confidently try a crowd or hybrid approach as it will not waste money. 

In essence, the approach - which we call \textit{hybrid shortest run} (HSR) -  works by considering the output of machine classifiers as a prior to the class probability for each item. 
This information is combined with an adaptive crowdsourcing algorithm that refines the class probability by polling the crowd on the (item,filter) pair that makes it more likely to reach a decision cheaply. 
In turn, classification information that is progressively obtained as we poll the crowd can be used to re-assess the performance of ML classifiers as well as create a model over the available classifiers for improved accuracy. 

While many adaptive algorithms have been proposed (e.g., \cite{kamar_crowdonly_2013,POMDP_2013,Nushi2015CrowdAP}),  to the best of our knowledge this paper is the first to discuss hybrid classification in screening contexts, which has unique requirements and opportunities since, as we will see, the heart of the problem lies in identifying the filters (e.g., the exclusion criteria in SLR) that i) are most selective (screen out a large proportion of papers) and ii) that crowd or machines can classify accurately (can correctly determine if the exclusion criterion applies to a paper). If we can do so, we can focus on getting crowd votes for these filters first, leaving a smaller set of items for the filters the crowd finds hard to classify correctly or that do not have high selectivity.

We believe this paper is also the first to take a "black box" approach to  hybrid classification: since we tackle finite pool problems, we do not assume we can develop an accurate, dedicated classifiers for each classification problem, although we may have available (possibly weak) classifiers from similar problems - in our SLR example these may be neural nets developed for other SLRs in the same or different fields or for other exclusion criteria. 
Rather, we identify how to leverage classifiers we are given and for which the accuracy for our task is unknown. 
We do this by a combination of (minimal) testing to filter out very poor classifiers  and by combining the remaining ones to set a prior probability for whether a filter applies to an item. This helps us determine the most promising (item,filter) pairs for which asking a vote to the crowd has the highest probability of reaching a decision cheaply and confidently. 

While it is not surprising that hybrid approaches have the potential to outperform human- or machine-only methods, the contribution we provide here lies in showing how we can cope with finite pools problems where potentially weak classifiers - and  even classifiers with unknown accuracy for the problem at hand - can be leveraged effectively, even in contexts characterized by very demanding requirements in terms of accuracy (as is the case for SLRs) and in the presence of difficult problems where also the crowd has low accuracy. 
HSR always estimates the characteristics of the problem and the classifiers, and identifies which are the items and filters for which it can draw accurate conclusions cheaply, leaving the rest to expert classification.

We also study how the effect of incorporating classifiers (and ensembles of classifiers) into a crowdsourcing algorithm varies based on the different characteristics of the problem at hand (such as different selectivity of filters, accuracy of classifiers, correlation among base classifiers, amount of data available for testing classifiers and for training ensembles) and show why and under which conditions \textit{ensembles} of classifiers do or do not provide benefits over "simply" using the best available classifier at hand. 

\input{sections/related}

\section{Model, Assumptions and Problem Statement} 
We next define the problem we aim at solving and the model, which captures a few broad assumptions we make.

Specifically, we assume to have in input a set $(I,F,C,L)$ where $I$ is the set of items to classify (in our SLR example, these are papers to screen), $F$ denotes the filters (paper exclusion criteria), $C$ represents the set of ML or H classifiers, and $L=k*FE + FI$ is a loss function, modeled as a linear combination of false exclusions FE and false inclusions FI, which may carry a different relative weight $k$.
The reason for the weight $k$ is that in many contexts the two kinds of errors have very different implications. 
This is for example the case when screening is performed to reduce the number of cases to be brought to human attention, such as potential credit card fraud, tweets/posts potentially linked to criminal activities, or SLRs: screening out an item is often more  ``costly" than erroneously bringing the item to human attention.  

The (possibly empty) set of given ML classifiers is assumed to be trained to receive an (item, filter) pair and output whether the filter applies to the item (e.g., receive a paper and an exclusion criterion and assess if the criterion applies to the paper). 
As mentioned, we do not discuss how to obtain classifiers (some examples are provided in the experiments section), as this is the subject of ample literature and it anyways depends on the problem domain. 
We do  \textit{not} necessarily assume that the classifiers are trained or even tested in the specific problem at hand (that is, for the specific filters and items under consideration),  and therefore in general we do not even know their accuracy. 


For human classifiers, we assume we have a set of generic workers of different and initially unknown accuracy and a set of high-accuracy but expensive experts. 



Each classifier $c = \{ cost, a(f), \rho(f,C) \} \in C$ is associated with the cost of asking one vote on an (item, filter) pair, with a filter-specific estimated accuracy (a 2x2 confusion matrix capturing probability of correct decisions for positive and negative labels), and with its \textit{error correlation} $\rho$ with other classifiers.
The error correlation among two classifiers is the probability that both make an error given that one of them makes an error.
Because we have no knowledge on the classifiers' accuracy, we conservatively model it as a uniform $Beta(1,1)$ distribution for both positively and negatively labeled items (which means we do not even assume classifiers are better than random). If we do have additional information, this can be incorporated in the  $Beta$. 
Consistently with crowdsourcing literature, we also assume that crowd workers opinions are independent (that is, they make independent errors, given an item, filter and true label). 

Each filter $f$ is characterized by  \textit{difficulty} $d_f$ and \textit{power} $\theta_f$.
Difficulty reflects how easy it is for crowd workers to classify items correctly on that filter. 
Following Whitehill \cite{whitehill2009whose}, we model difficulty as a  real number $d_f \in [0,+\infty)$ that, given an expected accuracy $\alpha_w$ of a worker $w$, skews the accuracy as follows: 
\begin{equation}
\alpha_{f,w} = 0.5+ (\alpha_w-0.5) \cdot e^{-d_f}
\end{equation} 
As the difficulty $d_f$ grows, $\alpha_{f,w}$ goes to 0.5, corresponding to random selection, which we consider to be the lowest accuracy level. 
The power $\theta_f$is simply the proportion of items to which  filter $f$ applies. Notice that we do not assume in this paper that power affects difficulty and we also do not assume different accuracies based on the true label of the items.
Both difficulty and power are unknown and we assume no prior knowledge on them.

Given this model and assumptions, our goal is to identify an algorithm that can  efficiently (in terms of cost) query the classifiers available and aggregate results while achieving the quality goals, stated in terms of loss as well as of the classical precision and recall measures.

\section{Baseline Algorithms}
We next describe baseline crowd and machine classification algorithms on which we base the construction of the hybrid approach. 
The crowd-based one is  \textit{shortest run} (SR)~\cite{Krivosheev_www18}, a recently developed algorithm that has been shown to perform well for multi-filter screening. We summarize SR here to make the paper self-contained. There are other reasons why SR is particularly suited for the hybrid approach we propose, and we get back to this in the following.

\subsection{Recap of Shortest Run}
The Short Adaptive  Multi-Run  algorithm (Shortest run, SR for short) identifies and continuously updates an individual strategy for each item to be screened by identifying the shortest path to decision. 
In a nutshell, the idea borrows concepts from Partially Observable Markov Decision Processes (similarly to \cite{kamar_crowdonly_2013,POMDP_2013}), by assessing based on the current state (that is, the knowledge accumulated thus far at the \textit{global} level - power and difficulty  for each filter - and at the \textit{local} level - the votes on each specific item) which are the (item, filter) pairs to submit to the crowd for testing because they would be more likely to quickly (cheaply) lead to an accurate decision.  SR may also decide to quit trying the crowd classification approach on an item if it believes that it would be too expensive to reach an accurate decision.  

SR proceeds by performing what authors call a \textit{baseline run} where a small set of items (usually 50) is screened with the standard approach of asking for labels on all filters (typically, 5 labels per item and filter) and classifying using a variant of Expectation Maximization. 
The result of the run is information on the power $\theta_f$ and difficulty $d_f$ (and, correspondingly, workers' accuracy $\alpha_f$) for each filter. 

This information is used to obtain a prior estimate on the probability that the \textit{next} vote for each (item i ,filter f) will be a vote to screen out the item (that is, a vote that the filter applies to the item). We get an out vote if the item is out (with probability $\theta_f$) and the worker answers correctly, or if the item is in (with probability $1-\theta_f$ and the worker answers incorrectly:

\begin{equation}
P(v_{i,f}=OUT) = \alpha_f \cdot \theta_f + (1- \alpha_f)(1-\theta_f)
\end{equation}

In addition, by applying Bayes, we know the probability of a filter $f$ screening (i.e., $P(i \in OUT_f)$) or  not screening ($P(i \in IN_f)$) an item $i$, once we obtain a label for the $(i,f)$ pair.

\begin{equation}\label{eq:bayespin}
\begin{split}
P(i \in IN_f | v_{i,f})= \frac{P(v_{i,f} | i \in IN_f ) * (1-\theta_f)} {P(v_{i,f})}
\end{split}
\end{equation}

Because an item is screened out if at least one filter applies, then:

\begin{equation}
	\begin{aligned}
		P(i \in OUT) = 1 - \prod_{f \in F} P(i \in IN_f)
	\end{aligned}
	\label{formula:p-out}
\end{equation}

An item is classified as out if $P(i \in OUT)$ is greater than a threshold $\overline{P_{out}}$. 
These formulas allow SR to estimate both what the next vote can be for a pair $(i,f)$ and the impact  each vote has on  $P(i \in OUT)$.
These estimates are updated as votes come in, where in Formula \ref{eq:bayespin} the class probability, initially derived only from filter power and therefore equal for all unclassified items, is updated based on the votes obtained thus far for that item.

The above formulas can be easily extended to compute the probability that the next $n$ votes for an $(i,f)$ pair will be \textit{in} or  \textit{out} votes. 
SR can in particular estimate the minimal number of votes $N^{min}_{i,f}$ in one direction (in or out) it needs to reach a decision, and the probability $P^{min}_{i,f}$ of getting such votes. 
As $N^{min}_{i,f}$ grows and its probability shrinks, SR may decide that crowd classification cannot be done efficiently, and quits trying (the stopping criteria are based on threshold which can be set as discussed in \cite{Krivosheev_www18}). 
Intuitively, items that are left unclassified are essentially not filtered out, so they contribute to a higher loss, which has to be weighted against the price one would incur by insisting with the crowd. This in general is part of the same trade-off of how much users are willing to spend for unit of loss.

\section{Hybrid SR algorithm}


\begin{figure}[htb]
  \centering
  \begin{minipage}{.7\linewidth}

\begin{algorithm}[H]
\nl \KwIn{Input: $I$, $C$, $F$, $T$, $\overline{P_{out}},\overline{P_{in}}$, $Budget$}
\nl \KwOut{$CI=\{CI_{in},CI_{out} \}$ \textit{\#classified items}}
\nl $CLF \leftarrow$ \textit{Select classifiers based on T tests}\\
\nl $C_{meta} \leftarrow$ \textit{Ensemble CLF}\\
\nl $\textbf{\textit{Prior}},  \hat{\theta^0} \leftarrow$ \textit{run ensemble on I}\\
\nl $CI \leftarrow \{\}$, $UI \leftarrow I$,$k \leftarrow 0$\\ 
\nl $Votes^0, \hat{\alpha^0} \leftarrow$ \textit{\textbf{Baseline run }}\\
\nl \textit{\#SRuns iterations}\\ 
\nl \While {$UI \not= \varnothing$ \textbf{and} $Budget \not= \varnothing$} {
\nl 	$k \leftarrow k + 1$\\
\nl     \For{$i \in UI$}{
\nl     	$f(i) \leftarrow $ \textit{assign filter on i | \textbf{Prior}, Votes}\\
\nl         \textit{do check stop condition on i | \textbf{Prior}, Votes}\\  }
\nl    $I^k \leftarrow$ \textit{items with highest prob of classification}\\
\nl    \For{$i \in I^k$} {
\nl    	$v_{i,f}^{k} \leftarrow$ \textit{collect a crowd vote on $f(i)$}\\
\nl		$Votes \leftarrow Votes \cup \{v_{i,f}^{k}\}$\\
\nl        \If{$P(i\in IN|Votes^{k}_{i}, \textbf{\textit{Prior}}) > \overline{P_{in}}$} {
\nl        	$CI_{in} \leftarrow CI_{in} \cup \{i \}$\\
\nl            $UI \leftarrow UI - \{i \}$\\
        }
\nl        \If{$P(i\in OUT|Votes^{k}_{i},\textbf{\textit{Prior}}) > \overline{P_{out}}$}{
\nl        	$CI_{out} \leftarrow CI_{out} \cup \{i \}$\\
\nl         $UI \leftarrow UI - \{i \}$\\
        }
    }
\nl    $\hat{\theta^k} \leftarrow$ update power\\
 }
\nl $CI^{difficult\_items} \leftarrow$ label $UI$ as \textit{"IN items"}\\
\nl \textbf{return} $CI, CI^{difficult\_items}$     

\caption{\textbf{Hybrid SR Algorithm}}
	\hrulefill
	\label{alg:hsm_runs}
    
\end{algorithm}
  \end{minipage}
\end{figure}

While there is ample literature on how to generate a ``good" ensemble of classifiers \cite{Rokach2010Ensemble,Dietterich_ensemble}, the prior art on the general problem of combining an existing set of ML classifiers $C =\{c\}$ is relatively less abundant \cite{Dzeroski2004Stacking}. 
We base the hybrid machine-crowd classification strategy on modifying SR.
In the following, we first motivate why we start from SR and then introduce hybrid shortest run (HSR) by presenting, at each step, first the intuition and then the formalization.

The reason for starting from SR is that i) it was designed for multi-filter screening and has shown to perform better than baseline algorithms for crowd multi-predicate classification, ii) it has a per paper and per item probabilistic model that can leverage prior knowledge on items and filters, and ML classifiers can provide such knowledge, and iii) the algorithm can adapt to work with different test items $T$ of different sizes. 
This is important, as test items can help us filter out ML classifiers with an expected accuracy lower than a threshold $\overline{a}$ (to be tuned as discussed later), and the more extensive set of crowd-classified items from the baseline can be used to a) assess independence among ML classifiers (which is necessary if we want to pool votes from ML classifiers with simple algorithms such as majority voting or Naive Bayes), and b) build an ensemble model out of the ML classifiers where the output of each ML classifier is a feature. 
This is similar to \textit{stacking}~\cite{Dzeroski2004Stacking} although we do not apply it over the training data used to build the individual classifiers (which we do not assume to have), but rather on the labels as estimated by the hybrid classification.

The Hybrid SR pseudo code is presented by Algorithm \ref{alg:hsm_runs}. The first step consists in filtering out classifiers that do not perform well for the case at hand. Specifically, we want to retain a classifier if we are confident it has an accuracy that is better than random.
In principle, every contribution that is better than random helps, especially if it is independent from the other classifiers and if we weigh the vote by the classifier accuracy.

As commonly done in crowdsourcing, we assume we have access to a high accuracy test dataset $T$ (obtained at a cost $C_T=ec*\left\vert{T}\right\vert$, that increases linearly by a factor $ec$ with the number of test examples, where $ec$ represents the cost for an ideal, expert screening). 
We then assign a prior probability distribution to each ML classifier and for each filter (ML classifiers may perform differently for different filters), reflecting our belief of how well the classifier performs on the problem at hand. 
As mentioned, in absence of additional information, we assume a 
$Beta(1,1)$ uniform prior for a given filter.
Other choices of prior are possible, assuming more likely that classifiers will have an accuracy above 0.5 and that accuracies close to 0 or 1 are very unlikely (for example, a Beta (3,2) has such characteristics). If we have a non negligible number of test cases the impact of this choice on the performance of the algorithm is relatively small. 
In the following for simplicity we assume a Beta (1,1) prior for all filters.

ML classifiers are tested with the gold dataset $T$ (line 3), resulting in posterior probability $Beta(1+correct\_answers, 1+failed\_answers)$ which has a known distribution. 
We can, in particular, retain classifiers whose probability of being better than random is greater than a selection threshold $sc$, and this can be easily computed from the Beta distribution. 
Then we build an ensemble of ML classifiers and run it on a whole unclassified pool of items $UI$ (lines 4,5).

Consistently with SR, we then perform a baseline run on $B$ items (on all filters), both to estimate crowd accuracy on each filters and to get data for the next step (line 7).

This information is used to inform a prior $p^{mc}_{i,f}$ for each item and filter in Equation~\ref{eq:bayespin}. 
In other words, while SR takes as prior for each filter $f$ the proportion $1-\theta_f$ of items classified as ``in" for $f$ and computes an overall prior, here we use the baseline ensemble output for each prior $p^{mc}_{i,f}$, where the class probability is the confidence that the filter applies.
Equation~\ref{eq:bayespin} therefore becomes

\begin{equation}\label{eq:bayespin_ml}
\begin{split}
P(i \in IN_f | v_{i,f})= \frac{P(v_{i,f} | i \in IN_f ) * (1-p^{mc}_{i,f})} {P(v_{i,f})}
\end{split}
\end{equation}

Notice that the prior has three effects: 1) it concurs to determine the classification probability; 2) it also affects which (item, filter) pairs we pick next for obtaining the crowd vote, since Equation~\ref{eq:bayespin} also affects $N^{min}_{i,f}$; 3) it has an impact on the stop condition (line 13), i. e., whether to stop to iterate over an item due to its classification difficulty or to continue. Once the $(i,f)$ pairs to query next have determined, HSR proceeds to obtain the vote and iterates as per the SR algorithm (line 16). 

We now present a series of experiments that, besides assessing the validity of HSR, help us understand its robustness as  parameters of the problem and of the algorithm change. Besides comparing it with crowd-only classification, we also compare with ML-only classification. For this baseline comparison we leverage a Naive Bayes classifier, where machines are considered to be independent and where their votes are weighted by the estimated classifiers accuracy $A_c = \{a_c\}$ (by means of tests $T$). The ensemble is therefore defined as follows:

\begin{equation}\label{eq:ensemble_naive_bayes}
\begin{split}
Class(l \in \{0,1\}) \propto \prod_{c \in C} a^{\delta(l, l_c)}_c \cdot (1-a_c)^{1-\delta(l, l_c)}
\end{split}
\end{equation}

where $	\delta(l,l_c) = \begin{cases} 1 & l = l_c\\0 & l \neq l_c\end{cases}$ is the Kronecker delta function\footnote{https://en.wikipedia.org/wiki/Kronecker\_delta}, and $l_c$ is a label from a classifier $c$.

\section{Analysis and Experiments}
The approach to analysis is based on both simulations and crowdsourcing experiments. The simulations are interesting because they allow us to understand the behavior of the approach under very different conditions, and assess the impact of the variation of each parameter, be it a parameter that describes the nature of the problem (such as the filters' power and difficulty) or a parameter of the algorithm, such as the classification threshold $\overline{P_{out}}$.

The crowdsourcing experiments allow us to get actual values of parameters for real scenarios and assess validity of results in that context - besides understanding a set of nuances important in the setup of crowdsourcing tasks.
We also leverage the data to build classifiers using commonly available techniques and use their accuracies to get a feel for the kind of accuracy we can achieve for the problem at hand.

\textit{Metrics.} 
In the experiments there are three metrics we want to assess: 
\begin{enumerate}
\item The \textit{loss}, computed as defined earlier, that quantifies our error weighted by how ``severe" false exclusions are considered (we can similarly use F$_\beta$ for this, though we choose loss in line with previous literature and also as it is more intuitive).
\item The \textit{recall}, which is as usual defined as \textit{true inclusions / (true inclusions + false exclusions)}
\item The relative cost of crowdsourcing, defined as the \textit{price ratio} of the cost of crowd classification versus expert classification. 
\end{enumerate}


\begin{figure*}[hbt]
     \subfloat[\label{fig:loss-price} ]{%
		\includegraphics[width=.333\textwidth]{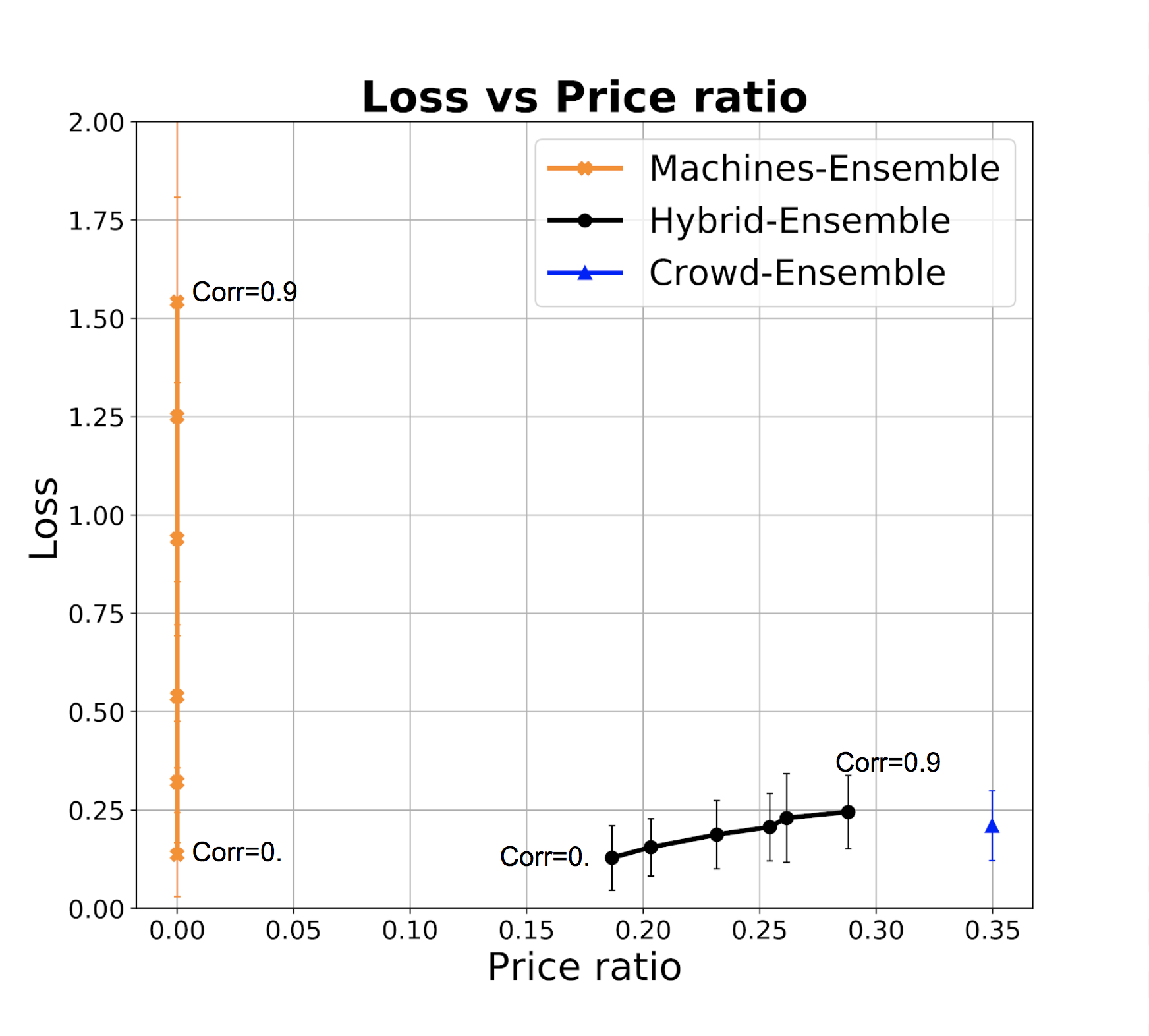}
     }
     \subfloat[\label{fig:recall_vs_pr}]{%
       \includegraphics[width=.333\textwidth]{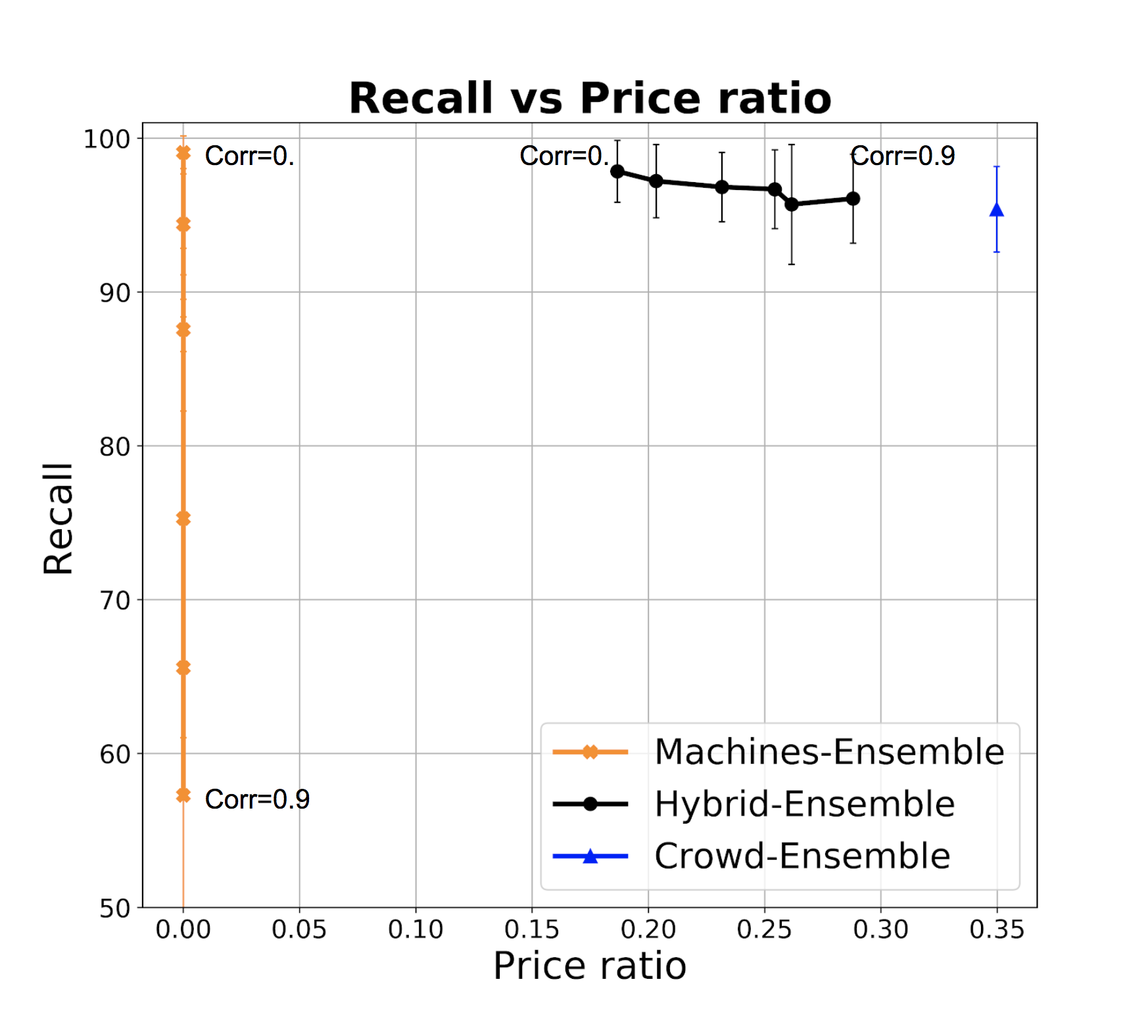}
     }     
     \subfloat[\label{fig:pr_vs_expertcost}]{%
       \includegraphics[width=.333\textwidth]{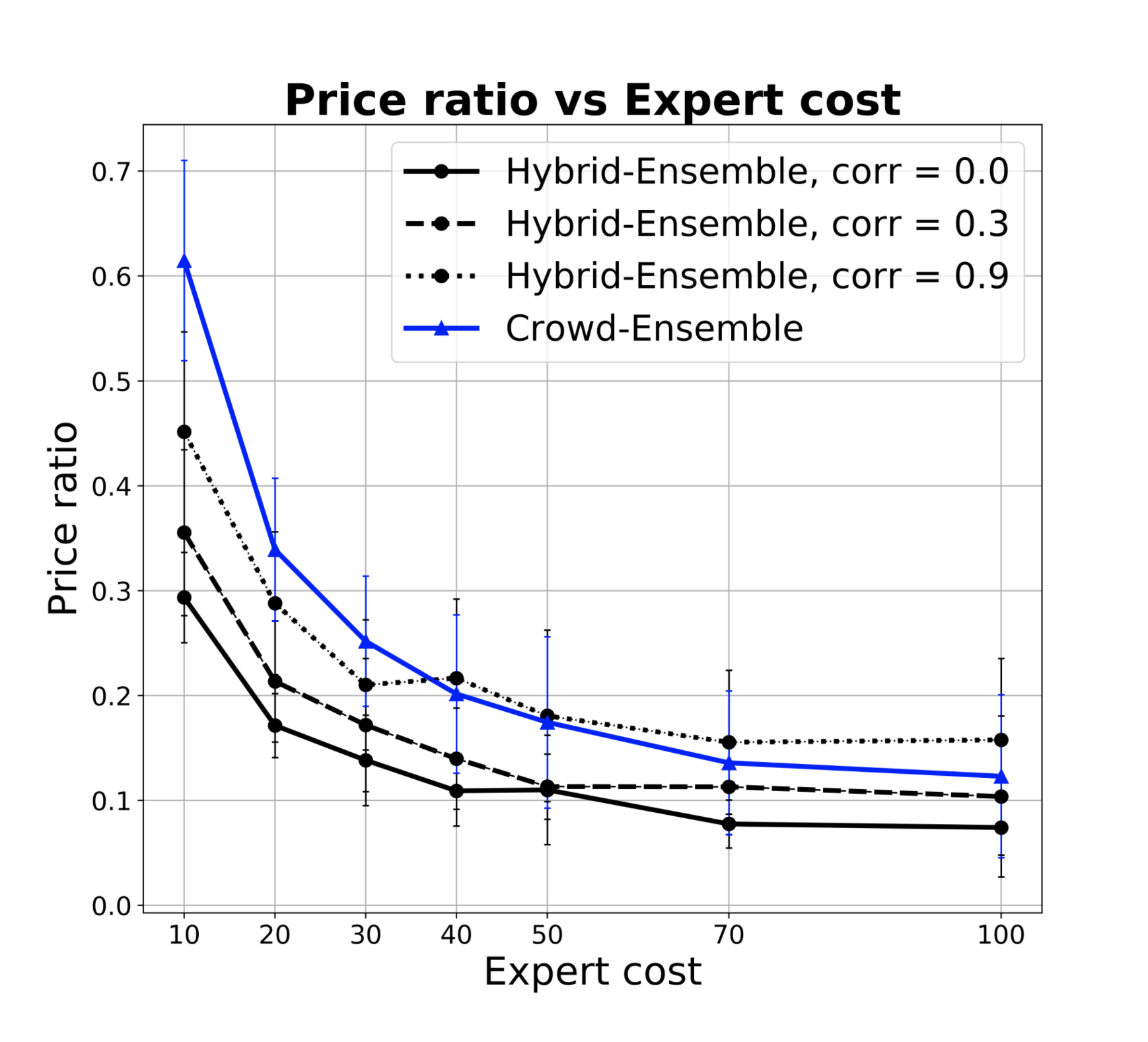}
     }
     
     \subfloat[\label{fig:recall_vs_theta}]{%
       \includegraphics[width=.333\textwidth]{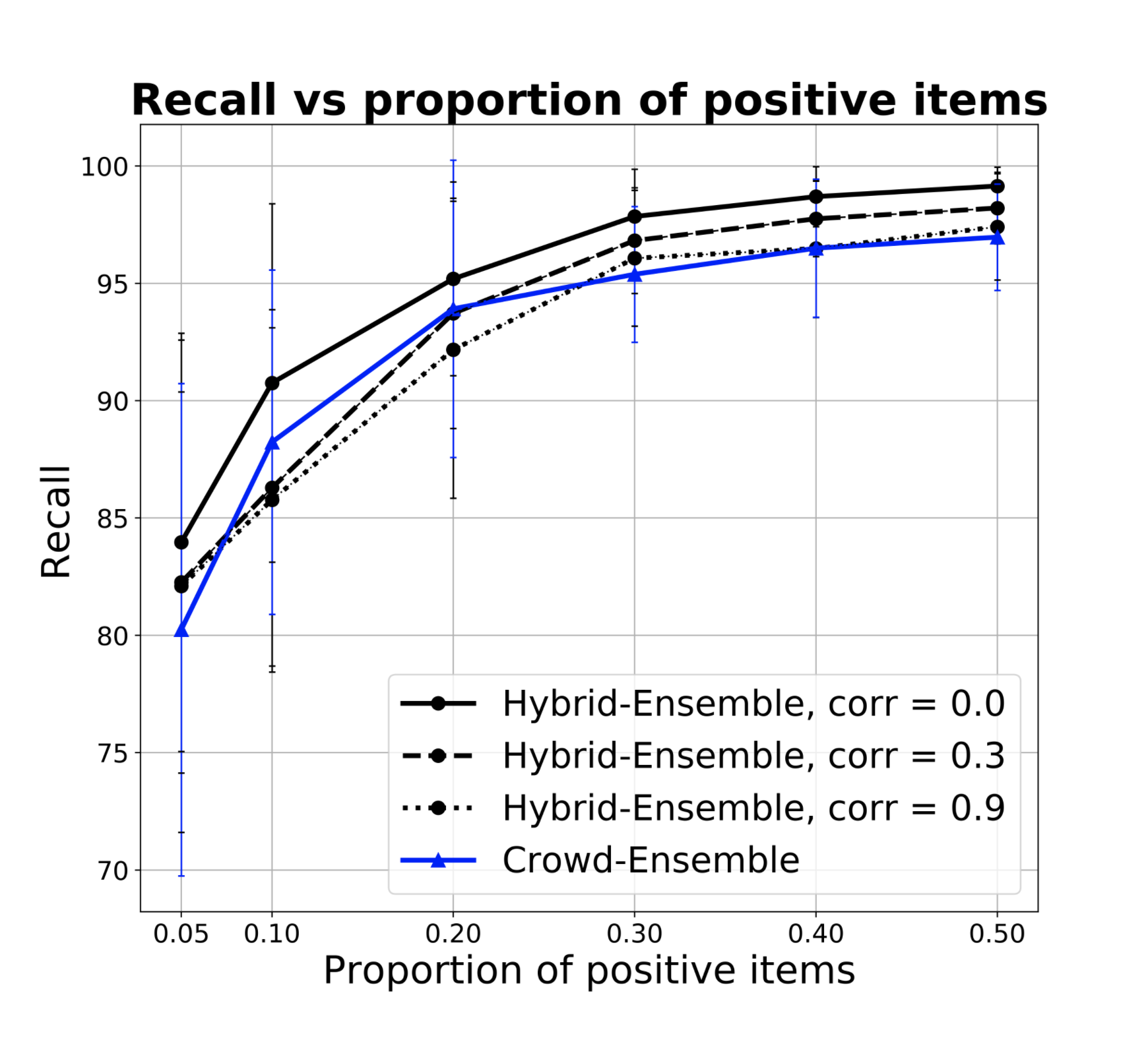}
     }
     \subfloat[\label{fig:pr_vs_theta}]{%
       \includegraphics[width=.333\textwidth]{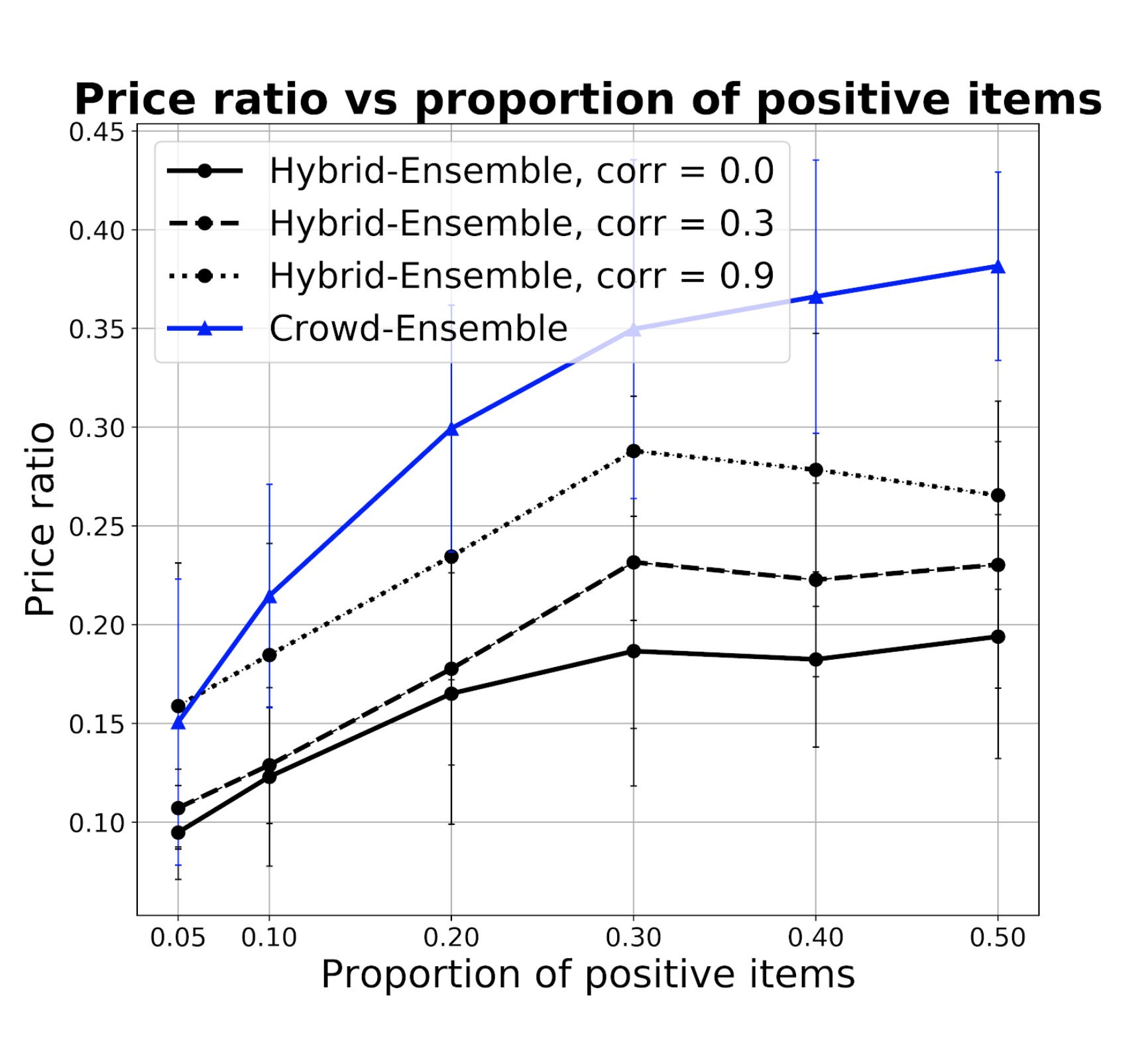}
     }
     \subfloat[\label{fig:vary_filter_params}]{%
       \includegraphics[width=.333\textwidth]{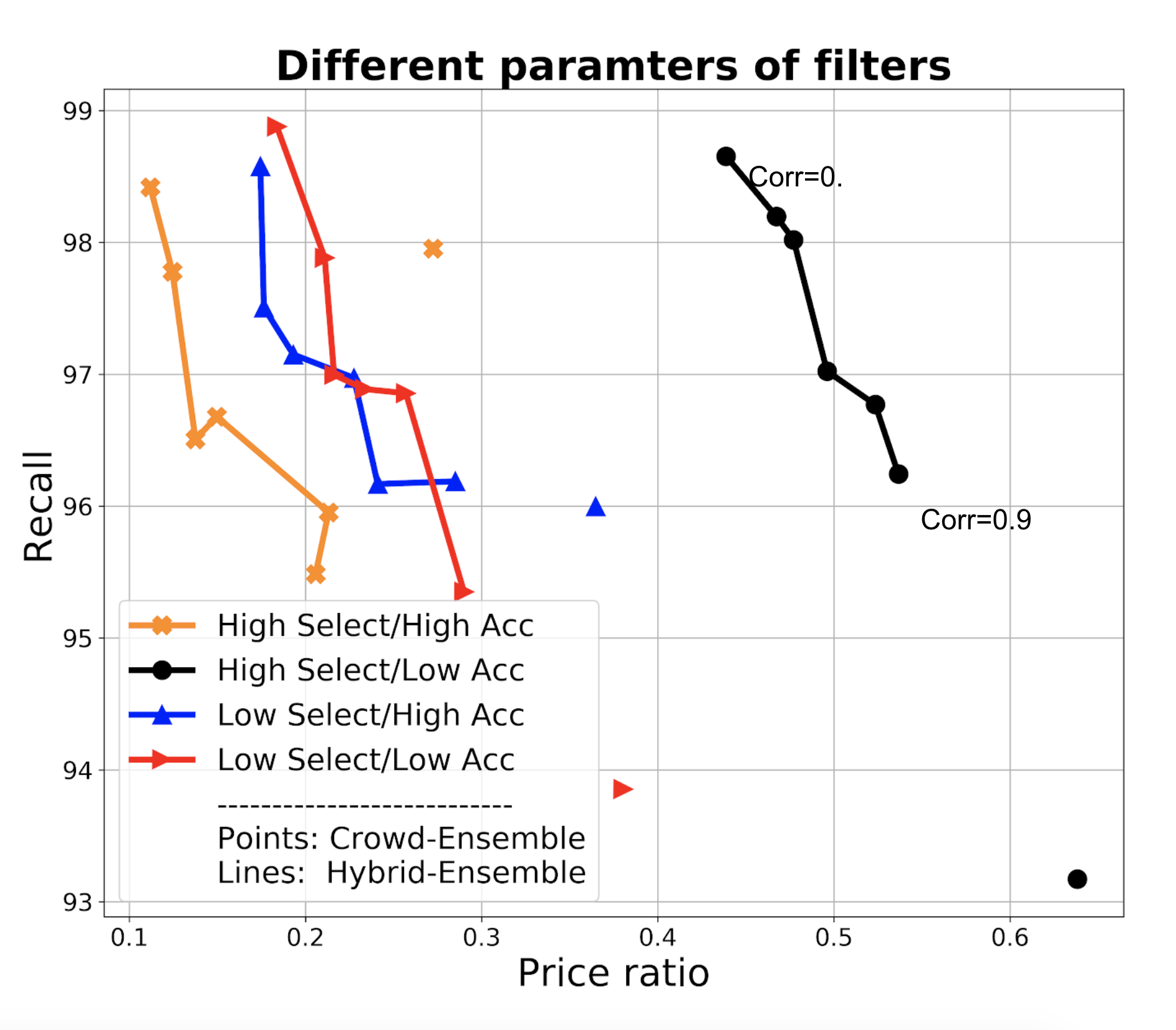}
     }           
    \label{fig:simulations}
\caption{Analysis of the hybrid algorithm as key parameters vary. Settings and description are provided in the text. Crowd-Ensemble refers to the SR algorithm, while Hybrid-ensemble is HSR.}
\end{figure*}

The rationale for using price ratio as a metric when comparing algorithms is as follows: we consider that the price of crowdsourcing is not just represented by the total number $CV$ of crowd votes asked: 
In fact, for each false inclusion $FI$ (for each item the crowd fails to screen out) we need to incur in the expert screening cost, as we are leaving the item on the expert's desk. 
If we fail to account for this cost we overestimate the performance of the algorithm (incidentally, this is true for SR as well, although this is not discussed in the original paper and price ratio is not used there as a metric). 
We therefore compute the crowd cost as $CC = CV + FI *ec$\footnote{There are cases where the ``expert cost" cannot be easily computed as the cost of expert labor: for example, a quasi-exact medical screening test may present practical challenges and even risks. In these cases, either domain experts can map these considerations into an ``expert cost" metric, or the analysis needs to focus on precision, not price ratio.} and compare it with what the total expert cost $EC = \vert I \vert \cdot ec$ (recall that $I$ denotes the set of items to be classified) would be if experts did all the classification.  The price ratio is then $CC / EC$.
Notice that considering price ratio and loss also incorporates precision: lack of precision (false inclusions) means higher loss and higher classification cost (higher price ratio). For this reason we do not plot precision in the following.
To describe the results, we first describe the baseline parameter settings for the experiments, and then show how results changes as we vary each parameter.

\textit{Baseline experiment.} 
The baseline experiment settings consists of a problem where we screen items via 4 filters, and where 30\% of the items survive the screening - if we classify them correctly.
We simulate 10 ML classifiers with accuracy randomly selected from a 0.5-0.95 range and on which the algorithm assumes no prior knowledge. 
We screen them with T=50 tests, work the math with the Beta distribution as described earlier, and keep the classifiers with 0.95 probability of having an accuracy greater than 0.5. The classification threshold $\overline{P_{out}}$ is set at 0.99, the loss ratio $k$ in the loss function is set to 10, and the expert classification cost $ec$ is set to 20 times the crowd label cost. This latter value is estimated from~\cite{Mortensen2016crowd} for the case of SLRs. 
Notice that expert screening cost is \textit{per item}, while the unit label cost is \textit{per label} on an (item,filter) pair. 
We stress again that users only have to set parameters corresponding to their requirements in terms of loss function, and the algorithm works out the rest, including the expected price ratios and recalls corresponding to given thresholds $\overline{P_{out}}$.

In Figure~\ref{fig:loss-price} we compare the results of applying machine only (ensembled with Naive Bayes), crowd (with SR), and HSR to simulations of classifications for 1000 items. 
The charts plot the mean loss by price ratio per item, obtained by averaging the results of 50 iterations. 
Vertical bars denote standard deviation. 
The different dots of the machines and hybrid curves denote different values for \textit{error} correlation among machines, simulated at 0, 0.2, 0.3, 0.5, 0.7, 0.9. 
As we can see (Figure~\ref{fig:loss-price}), for a similar loss level, hybrid algorithms significantly outperform the crowd in terms of both price ratio and loss. 
Not surprisingly, price ratio and loss worsen as error correlation increases, as ML classifiers tend to agree when they make mistakes. 
However, even at high correlations we maintain performances that are superior to those of crowd-only. 
As we do not consider crowd correlation, for the crowd we just have a point in the chart.
ML-only classification is competitive only with independent and accurate classifiers, and their performance quickly deteriorates even at small level of error correlation, as we do not assume we have strong classifiers. 
In the case of SLRs, with expert accuracy in the 95-98\% range \citep{Krivosheev_www18,Mortensen2016crowd} and assuming that the items that survive the screening are in the 10-30\% range, the loss lies within 0.04 and 0.18 (with the settings of the experiment being closer to the 0.18 mark, slightly better than the result of HSR but a much higher cost).



Similarly, Figure \ref{fig:recall_vs_pr} shows the recall vs the price ratio. Recall is independent of the loss parameter $k$ and gives us a direct indication of our capacity to identify the items that pass the filters	 for a given monetary investment.
Notice that we disregard here the cost of obtaining the base classifiers, which, if they  are built from scratch for this specific SLR, needs to be factored in when estimating price and  assessing the best strategy. On the other hand, if built for the specific problem they are also likely to have higher accuracy. We do not discuss this further as again building the base classifier is orthogonal to our goal here. 
We also observe that recall is high "by design" in that we fix the threshold $\overline{P_{out}}$ for considering an item as screened. 

\textit{Effect of changes to the parameters.} 
We now explore how results changes as we change either parameters that describe the nature of the problem (e.g., filter power, crowd accuracy, or loss and expert cost ratios) and the behavior of the algorithm (e.g., decisions on how many test data to obtain, or thresholds to set). 
Here we focus on a few interesting variations, and refer the reader to our GitHub repository\footnote{https://github.com/Evgeneus/crowd-machine-collaboration-for-item-screening} for additional details, data and source code.  
Notice that HSR is \textit{adaptive}, that is, it estimates the parameters of the problem and changes its behavior accordingly. For this reason, we expect the recall to remain high, while price ratio may change. 

Fig \ref{fig:pr_vs_expertcost} plots how the price ratio improves with the expert cost $ec$. Recall (not plotted here) is essentially constant. For example, for a 0.5 correlation, it stays in the .96-.97 range for all $ec$. 
The lines flatten for high $ec$ and asymptotically reach the minimal price ratio, which is the proportion of false inclusions.

Figures \ref{fig:recall_vs_theta} and \ref{fig:pr_vs_theta} show instead how recall and price ratio change with the proportion of items that should pass the screening (positive items). Specifically, in the chart we show the behavior when the power of one of the four filters grows.
The figures show that the performances worsen in high power situations. This is because high powers mean a prior $P (i \in OUT)$ that is very close to the $\overline{P_{out}}$ threshold, and as such it is more sensitive to  errors from the crowd in the first few votes. This is not a problem of HSR per se but is inherited from SR. 
To correct it, it is sufficient to "tone down" the prior when the power estimate is too high. 
Indeed, we observed experimentally that this can be achieved by structurally underestimating power by approximately 20\% of its value. 

Finally, Figure~\ref{fig:vary_filter_params} shows how recall varies in the presence of one criteria that heavily differs from the other in terms of difficulty (accuracy) and power (selectivity of filters). As expected, recall remains constant because the algorithm adapts to the characteristics of the problem. Price ratio worsens as we go from an easy and powerful criteria to the most difficult case of low power, low accuracy filter where we have many incorrect votes, and even correct ones do not help us much because the screening power of this filter is low, which means we anyways need to query the other filters.  
HSR results are robust to variations in the number of tests and in the confidence thresholds for keeping a ML classifier (not shown). Assuming a set of classifiers in the 0.3-0.7 range instead of 0.5-0.95 cuts approximately in half the gain with respect to SR.




\textbf{Experimental data and experiment design.}
We experimented both by using existing datasets and by running crowdsourcing experiments.
In all cases, data and code for the algorithms are available at the same url [omitted], along with other charts describing variations of behaviors with parameter values which we skip here as we find them  less informative.



\begin{figure}[hbt]
	\centering
    \begin{minipage}{.8\linewidth}		\includegraphics[width=0.7\textwidth]{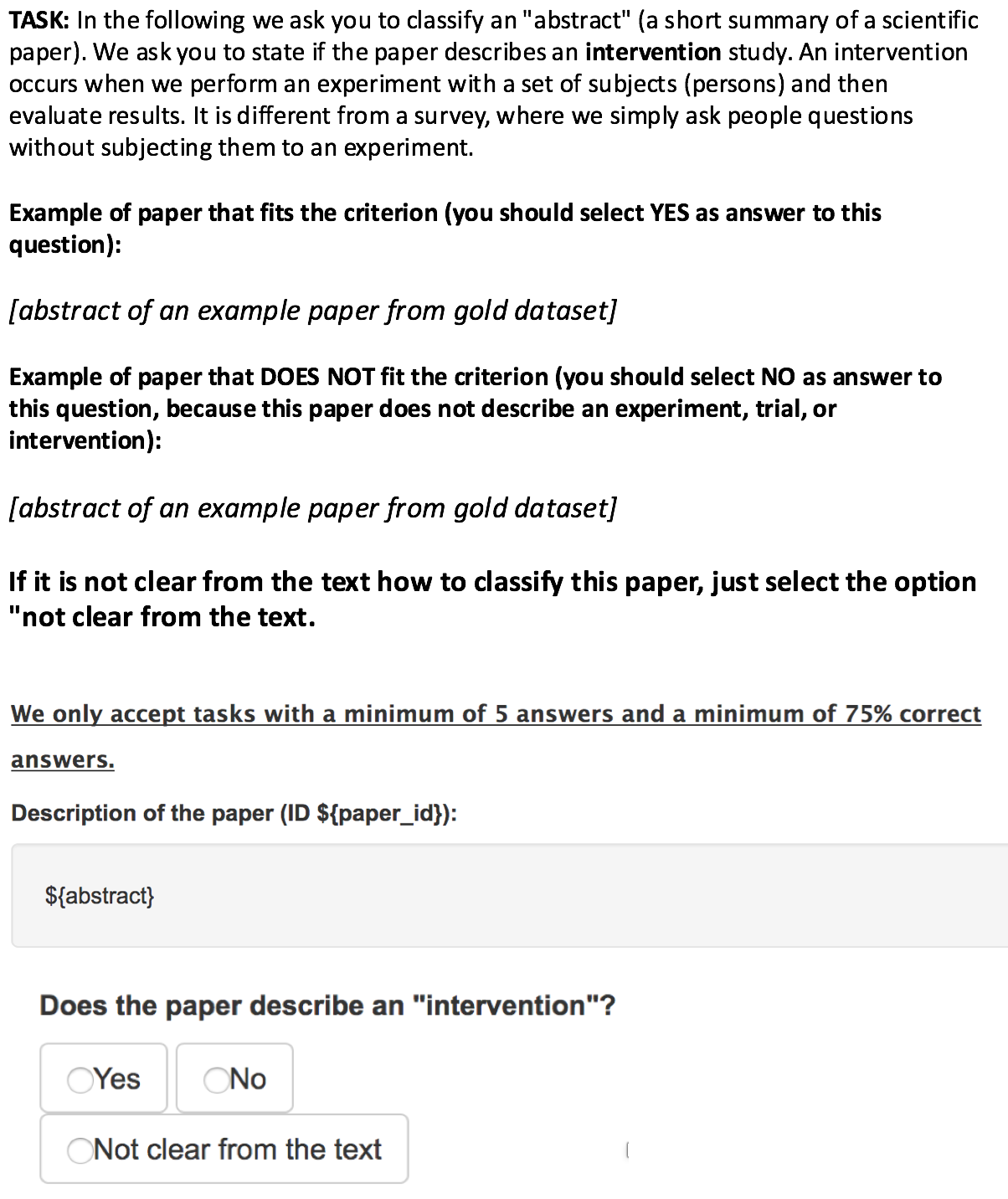}
		\caption{Classification task for SLR (settings borrowed from \cite{Krivosheev_www18})}   
\label{fig:task_design}
	\end{minipage}
\end{figure}

\begin{figure*}[hbt]
     \subfloat[\label{fig:recall-tests} ]{%
		\includegraphics[width=.383\textwidth]{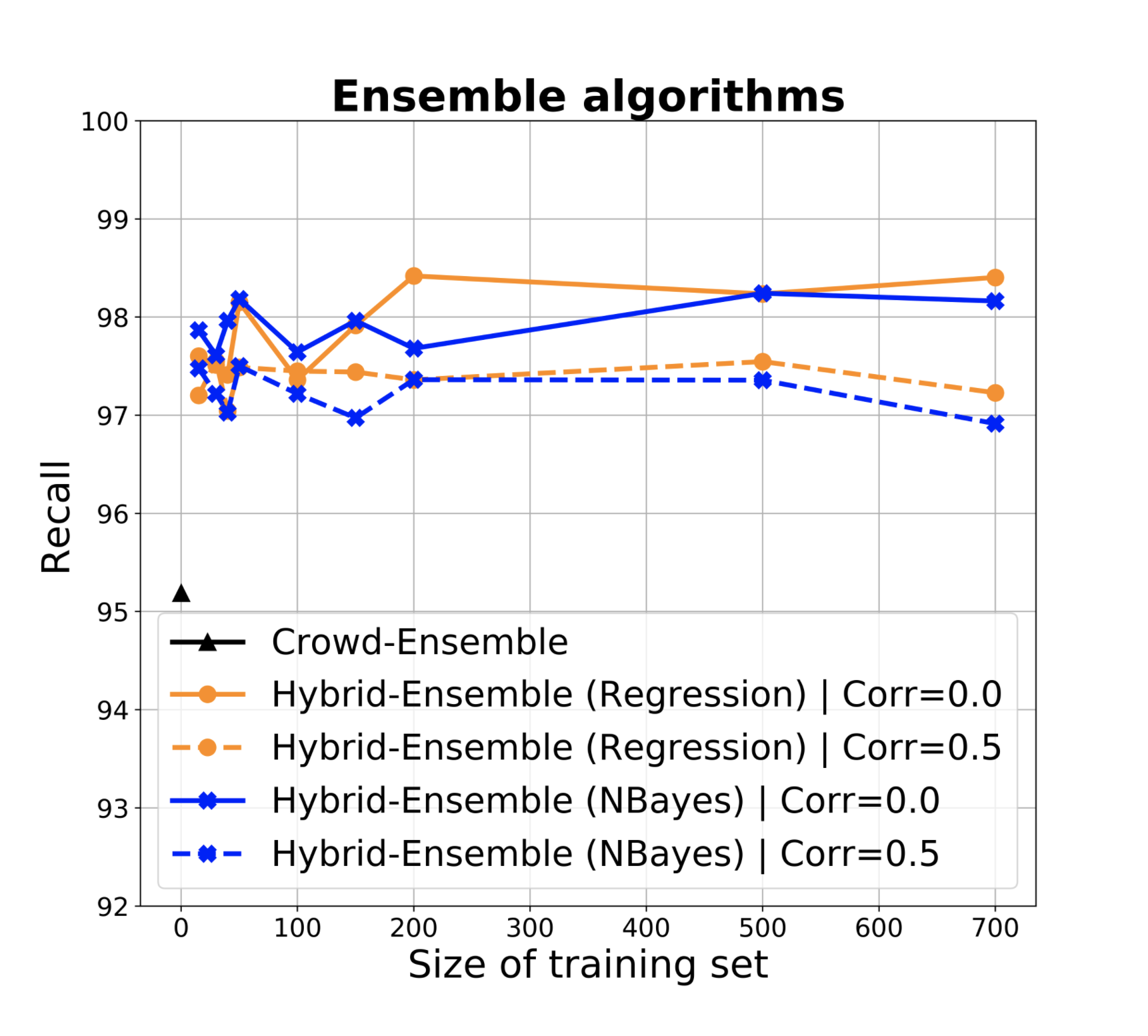}
     }
     \subfloat[\label{fig:recall-price-acc}]{%
       \includegraphics[width=.383\textwidth]{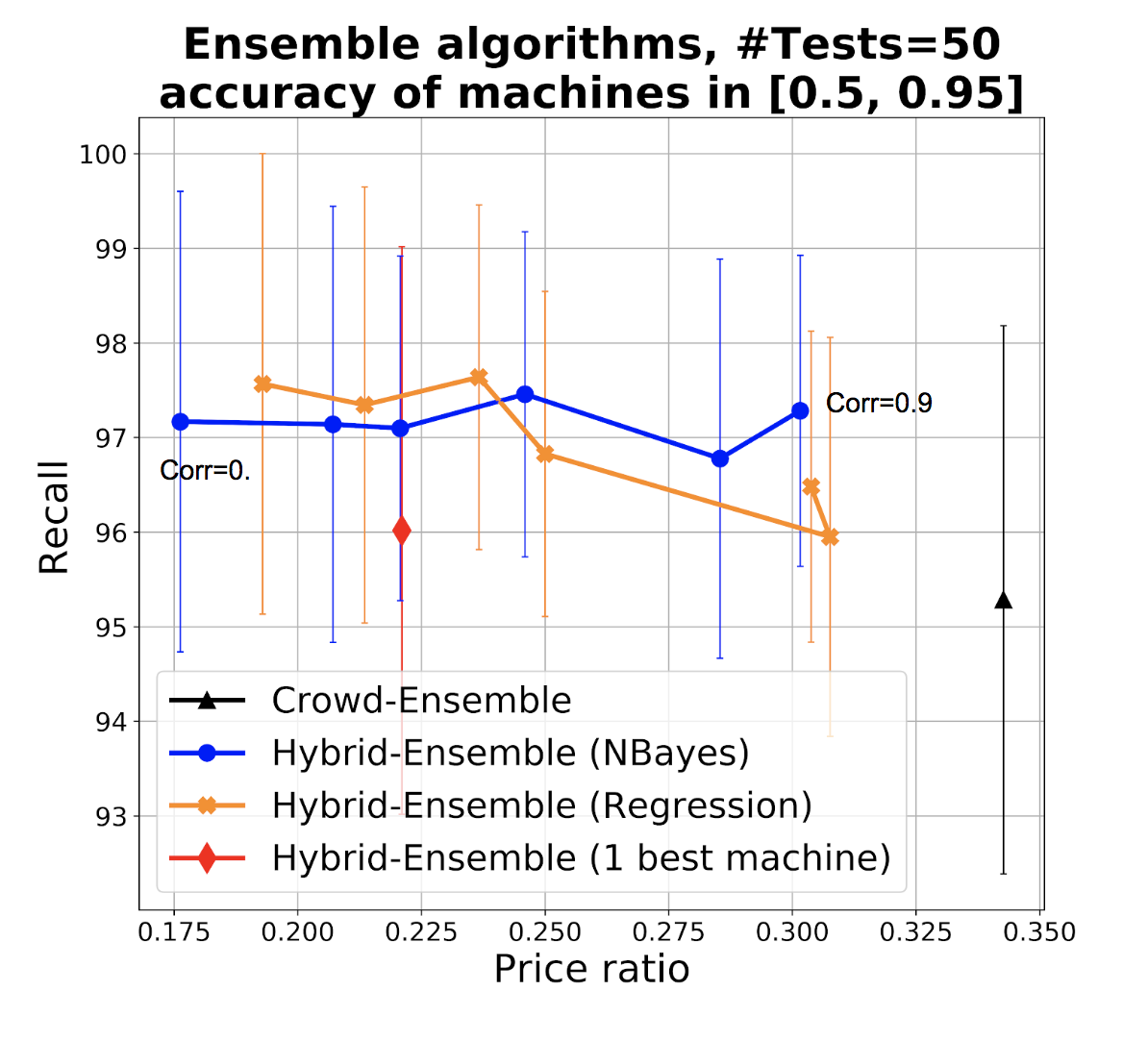}
     }    
     
     \subfloat[\label{fig:fig3c_ensemble_algo_acc05_075}]{%
       \includegraphics[width=.383\textwidth]{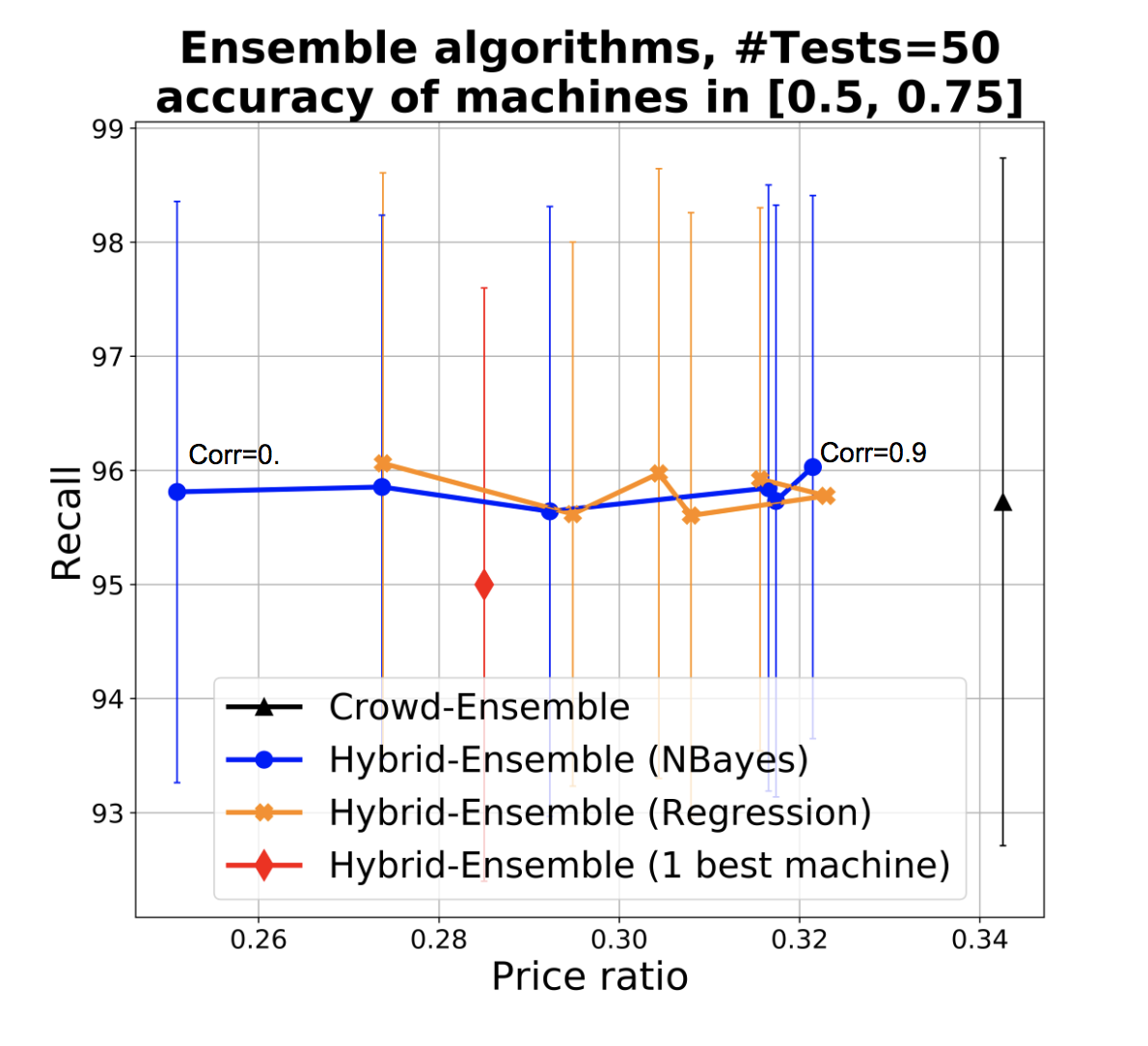}
     }     
     \subfloat[\label{fig:semi-real-experiment}]{%
       \includegraphics[width=.358\textwidth]{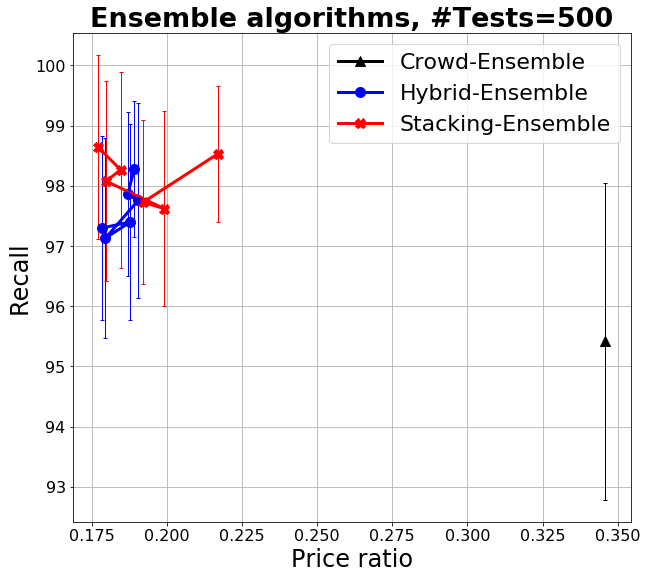}
     }
    \label{fig:simulations}
\caption{Experimental results for basic and stacking models. (a): recall for SR (black), basic HSR (blue) and HSR where classifiers are combined with regression (yellow) as the size of the training set used for the regression model varies; (b) and (c): recall vs price ratio for different distribution of classifiers accuracies; (d) results based on parameters obtained from AMT experiments. The dots in figs b,c,d present the correlation level of machine classifiers. Corr is [0, 0.2, 0.3, 0.5, 0.7, 0.9] from left to right. }
\label{ff}
\end{figure*}

We first took an existing SLR dataset by Wallace and colleagues~\cite{Mortensen2016crowd}. 
This includes over 20000 crowd votes over 4000 papers, with four filters. 
We also run a set of experiments on AMT and CrowdFlower, with different designs, but specifically borrowing the design from \cite{Krivosheev_www18} for a fair comparison, and eventually built our own platform on top of AMT for increased flexibility and for being able to plug in the adaptive algorithms into the task assignment logic (\textit{url omitted}). 
Specifically, the basic design included classifications of 100 papers for an SLR in the social informatics field, with three filters of different difficulty  (one of them very difficult, with a worker accuracy barely above 0.5). The filters were: "does the paper describe a study on 65+ older adults", "does the paper describe a study that uses technology", and "does the paper describe an \textit{intervention} type study" (the difficult one).
The task proceeded as depicted in Figure \ref{fig:task_design}. Workers with historical overall AMT accuracy over 70\% were invited to participate. A worker would see a description of one criterion, along with a positive and a negative example. They would then proceed to labeling papers based on that criterion. 
The choice of a per-filter approach (asking for labels on many items for one filter) as opposed to per-item, is because it takes time to properly explain and understand a criterion (for example, explaining what an intervention is may be tricky and requires the worker to go through positive and negative examples carefully).
We screen workers with two test questions and consider the labels from the remaining workers. After initial experiments to estimate the time taken by workers, we tuned parameters to arrive at a pay rate of 10USD/hour. We obtained 10 labels for each item and filter for a total of 3000 crowd votes from 147 workers.

We then built classifiers for each filter using a variety of techniques (from KNN to random forest, variations of naive Bayes and logistic regression and others, with different kinds of document representations) and different sizes of training data to get realistic information on  classifier accuracies and correlations. 
We obtained classifier accuracies in the 0.5-0.8 range, correlations in the 0.2-0.9 range, and crowd accuracy in the 0.55-0.8 range. 
In this case, classifiers were obtained by training on the same SLR data using 20\% of the data, where the training set is obtained via crowdsourcing, so the savings for the experiment refers to the remaining portion of the items to be screened.

We then use these parameters obtained from experiments  to fuel various experiments and repetition. 
We show the representative result in Figure ~\ref{ff}d where the blue line represent HSR. We ran experiments with different combinations of base classifiers, having different correlations among them as usual represented by dots in the line. 
Notice that the standard deviation is very high in proportion to the scale of the chart, though we can see the impact of the hybrid approach. Higher correlations (moving from top to bottom) lead to slightly lower recall. 
Incidentally, we also report that, consistently with expectations, errors made by HSR and by us (the experts, in this case) were about the same: half the time we disagreed with the crowd, either the crowd was right or the decision was questionable.

\textbf{Stacking classifiers.} 
In the experiments above we naively included ML classifiers regardless of their error correlation. However, if two or more classifiers predict the same class (and especially if they make the same errors) then it does not help to pool their predictions - worse than that, when they make mistakes they make so by consensus thereby increasing our confidence in the incorrect prediction.  
There are essentially two approaches to deal with this: the first is to filter out highly correlated classifiers. 
If we denote with $c^j_{err}$ the event corresponding to classifier $c^j$ making an error, then we look to exclude classifiers where $P(c^j_{err} | c^k_{err})$ is high, and particularly above 0.5, otherwise we reinforce the error of $c^k$.
This can be done again in the test phase similarly to what we do for estimating accuracy, for example again by assigning a Beta prior to this conditional probability. 
However, since errors might be rare, estimating error correlation requires more data points. If the initial (expert-provided) gold dataset is small, then we can proceed with HSR to get data points, initially with one or few ML classifiers, and then add more as we are confident of their low correlation.

However, we can also \textit{stack} the ML classifiers by learning a model, and let this model filter or downweight correlated classifiers. 
Figure~\ref{ff}(a-c) show the performance of a logistic regression model built over the base ML classifiers and leveraged in HSR, compared with base HSR as from the previous section (blue) and with crowd-only ensemble (SR). 
The plot assumes 5 ML classifiers with accuracies as before, and a varying size of training dataset. 
In theory, the regression naturally copes with correlation and in the presence of a sufficiently large training set, it identifies how to effectively combine classifiers.
As above, the training set can be progressively obtained via SR, and we can use base classifiers to identify a dataset that has higher possibility to be balanced, to get a better training set.

The charts show that in practice stacking with regression offers no improvement with respect to the naive aggregation of classifiers, and improvements are also limited with respect to taking the single best classifier in the set as opposed to an ensemble (but notice that SR takes us already to 0.95 recall, so improving from that is challenging). 

We found this result surprising, although consistent with the findings briefly mentioned by \cite{Dzeroski2004Stacking}, where however the problem is not discussed besides the mentioning of this observation. 
To get clarity on this matter we investigated it further. Specifically, besides experimenting with classifiers of varying accuracies and correlation, we engaged to understand if the lack of effect is due to the specific way we use the output of classifiers - that is, to set prior probabilities (which in turn determine how to query the crowd for that item based on the HSR algorithm) as opposed to directly take a classification decision.

\begin{figure*}[t]
     \subfloat[\label{fig:recall-tests} ]{%
		\includegraphics[width=.450\textwidth]{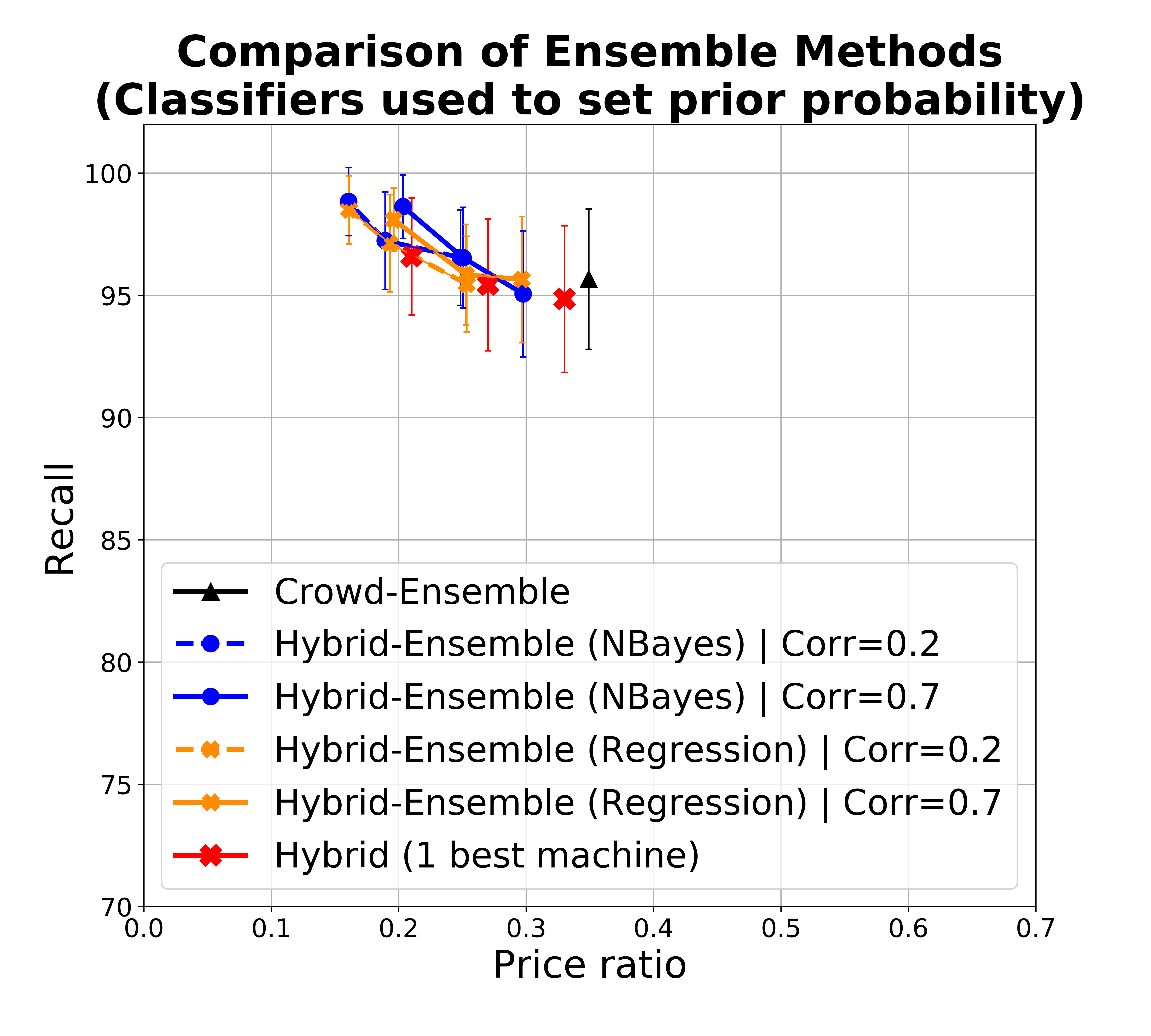}
     }
     \subfloat[\label{fig:recall-price-acc}]{%
       \includegraphics[width=.450\textwidth]{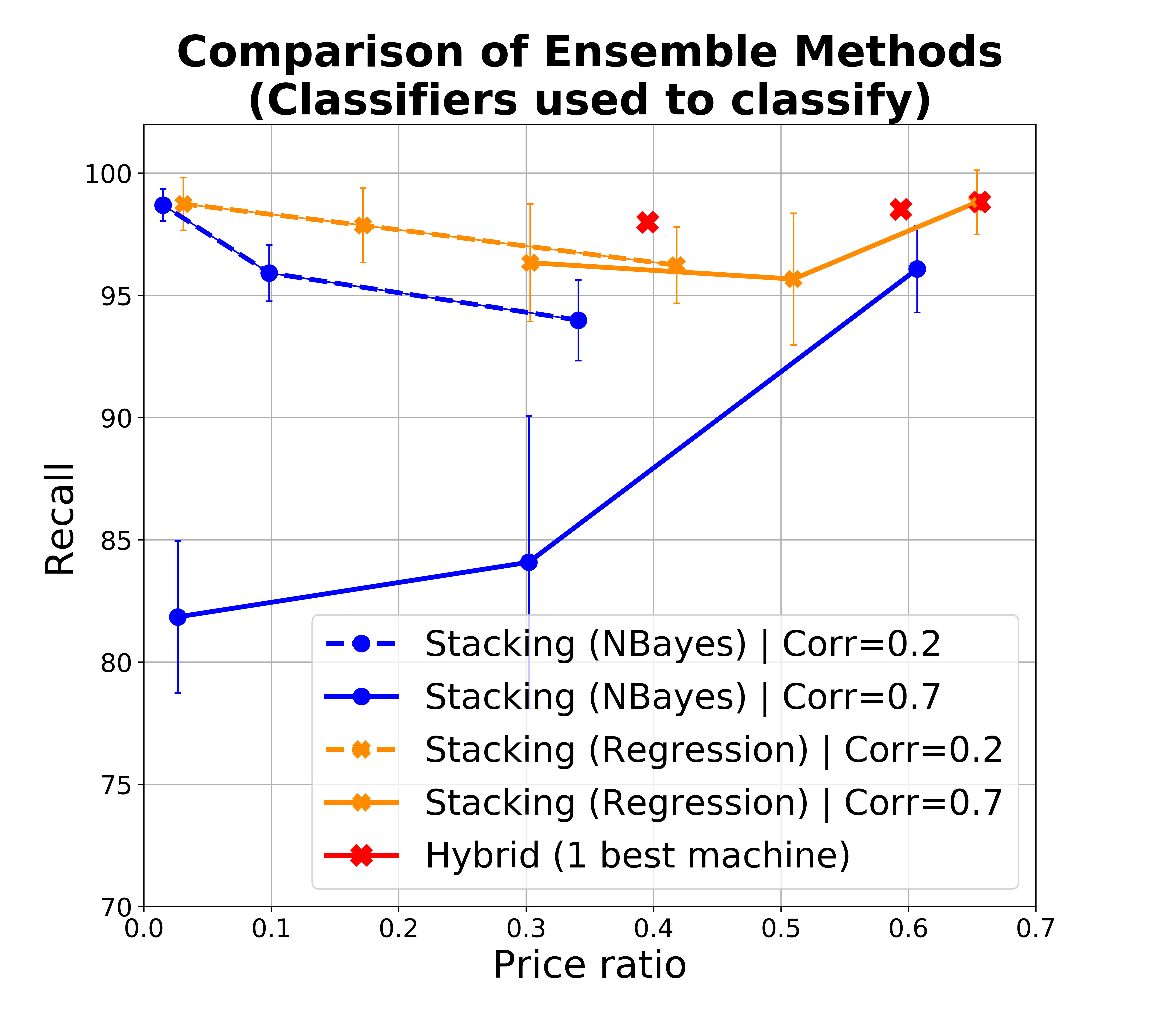}
     }   
    \label{fig:ensemble_simulations_b}
\caption{Comparison of NB and regression ensembles when used as prior (left) and to classify items directly (right). The dots in charts represent different accuracy range of machine classifiers ([0.65, 0.75], [0.75, 0.85], [0.85, 0.95]), increasing from  right to left. Ensembles have been trained over a dataset of 500 items.}
\label{fig:ensemble_simulations}
\end{figure*}

The results are shown in Figure \ref{fig:ensemble_simulations}. Figures a and b respectively show the performance of the ensembles when used as prior (Figure \ref{fig:ensemble_simulations}a) and when used directly to take final classification decisions, without resorting to the crowd (Figure \ref{fig:ensemble_simulations}b).
The different dots denote accuracy of the base classifiers, with recall and price ratio improving with accuracy as expected (with one interesting exception discussed later). The two figures are shown in the same scale to facilitate comparison. We adopt the price ratio  metric also for Figure \ref{fig:ensemble_simulations}b since classification errors mean that we incorrectly leave items for experts to classify, and this incurs in an unnecessary cost with respect to classifiers with perfect precision. 
Figure \ref{fig:ensemble_simulations}b shows that when we use the ensemble to make predictions, in low correlation conditions both Naive Bayes (NB) and regression ensembles are superior to the single best classifier especially in conditions of higher accuracy of the base classifiers (the difference in recall is small and within a standard deviation, while the benefit in terms of price ratio is high).
With high correlation, regression ensemble and best classifiers offer similar performances, while NB drops significantly in recall, our main target metric. This is not surprising since in conditions of high correlation, ensembles with NB essentially reinforce the errors. 
Indeed, as accuracy of base classifiers increases so does our confidence in their vote (and in the ensemble). Specifically, often this confidence will exceed our confidence threshold $\overline{P_{out}}$ for classifying items as out directly, which explains the drop in recall for the NB ensemble as accuracy increases.
In conditions of lower accuracy we take NB ensemble's predictions more cautiously, which explains the better recall (at the expense of price ratio).
The benefit of regression  however only manifests themselves once we have sufficient training data - in our experiments we started to observe a difference starting from training sets of 400 items. 

When we leverage ensembles as prior (Figure \ref{fig:ensemble_simulations}a) as per HSR algorithm, these differences smooth out considerably and the points in the charts become closer. In HSR we never classify based on machines only, as we ask at least one vote from the crowd, which helps avoid such drops in recall.
Ensembles improve over crowd-only approaches as discussed, even with weak classifiers, but because taking classifiers' results as prior is conservative and because classification threshold are high, recall errors from classifiers have a chance to be corrected by the crowd.

\section{Conclusions}
The main result of this paper is a method for multi-filter classification that combines ML and human classifiers to achieve high level of classification accuracy for unit of cost. 
We believe the main benefit lies in the ability of the algorithm to be robust to both the characteristics of the problem (such as  filters that are hard for the crowd to classify) and to weak ML classifiers. 
In both cases, the algorithm works out to leverage the filter, items, and ML classifiers (over specific filters and items) and builds models of classifiers to make the most of what can be screened automatically or with the crowd, leaving the rest for other classification methods (such as expert classification) without compromising on quality. 
Furthermore, the experimental results also helps us understand, in case we do have to build classifiers because we cannot transfer them from similar learning problems, what is the accuracy we should look for to get a benefit from hybrid classification.
While we focused on hybrid algorithms we also uncovered improvements to multi-filter crowd classification algorithms such as the opportunity to smooth the prior in high-power filter scenarios. 

Although the evaluation has focused on SLR we see the approach as applicable to many multi-filter screening problems.
An extensive evaluation of the limits of the approach and the kinds of problems to which it applies is part of our current work. 
In addition, we see active learning for ensembles in finite pool contexts such as the one described here as an interesting research problem, which is complicated by the likely presence of highly imbalanced datasets and by needs that are often conflicting, that of leveraging the crowd efficiently to screen items out but to also get training data classified as in. Another interesting thread of analysis is to see if crowd workers can, besides labeling (item,filter) pairs,  actually identify features that machines can then leverage for the problem at hand.

Finally, an interesting problem is the extent to which learning can be transfered across finite pool problems of the same kind (for example, different SLRs), especially in terms of identifying general approaches and methodologies that can be adopted in different domains.

%% file: sections/related.tex
\section{Related Work}

\subsection{Crowd Classification}

This paper researches a field that is abundant in prior art. 
The problem of crowd classification has been studied for centuries, at least since the early work of the Marquis de Condorcet who, trying to make a case for democracy, proposed his \textit{Jury Theorem}\footnote{http://www.stat.berkeley.edu/~mossel/teach/ SocialChoiceNetworks10/ScribeAug31.pdf}, stating that if each juror in a jury (or each citizen, when taking a binary decision) has an error rate lower than 0.5 and if \textit{guilty} vs \textit{innocent} votes are independent, larger juries reach more accurate results, and approach perfection as the jury grows. 

More recently, with the opportunity offered by crowdsourcing on the one hand and the need for large labeled dataset to support ML tasks on the other, many researchers from the AI, database, and human computation communities investigated the problem in detail, both to identify algorithms and to study theoretical bounds. Initial algorithms were based on variations of (weighted) majority voting, where votes are counted differently based on the estimated worker's accuracy.
Dawid and Skene \cite{DawidSkene_Confusion} in the late nineties started a thread of research based on modeling and iteratively estimating workers' accuracy and class labels, by adopting variants of Expectation Maximization \cite{EM_1977}.
From there, many important and useful extensions have been proposed by Whitehill \cite{whitehill2009whose}, Dong et al. \cite{Dong2013}, Li et al. \cite{Li_error_13},  Liu et al. \cite{LiuWang_truelabel,liu2013scoring} and many others. 
Relatively recent approaches based on spectral methods \cite{karger2011iterative} and maximum entropy \cite{Zhou_spectral_13} have also been proposed, and belief propagation has  been  recently shown  \cite{Ok_belief_13} to be optimal under certain assumptions. 

Prior art has also discussed how to continuously assess the estimated value of asking for more votes from the crowd versus taking a classification decision with the votes as available. 
The approach is based on modeling the problem in terms of partially observable Markov Decision Processes ~\cite{kamar_crowdonly_2013,POMDP_2013}, where policies are progressively learned as votes become available. 

\subsection{Crowd Classification in multi-predicate problems}
The set of classification problems we address here is that of \textit{finite pool} classification \cite{Nguyen2015}: we work on a finite set of items, and the classification criteria may be unique to the problem. SLRs are one case of such problems that has been thoroughly investigated by many researchers in recent years \cite{Mortensen2016crowd,Krivosheev_hcomp17,Nguyen-hcomp15,Sun16-hcomp}.


Mortensen and colleagues crowdsourced paper screening~\cite{Mortensen2016crowd} in four literature reviews, each with several criteria. They explore feasibility and costs of crowdsourcing, and they address the problem by measuring workers agreement.
Furthermore, they include a set of very interesting observations related to the importance of task design, as well as on the high degree of variability in workers's agreement from paper to paper and criteria to criteria (Fleiss' Kappa ranging from 0.5 to -0.03). 


Krivosheev and colleagues~\cite{Krivosheev_hcomp17} also present a model and strategies for crowdsourcing SLR. An interesting aspect of the model and approach here is that the authors model cost and loss (error) resulting from crowdsourcing task, attempt to estimate them at the start, and provide authors with a price/error trade-off that can be used to decide how much to invest in the task.  
They extend the work to the multi-filter screening case \cite{Krivosheev_www18}, also by introducing algorithms that efficiently query the crowd by focusing first on criteria that are more promising in terms of screening items correctly. 

We borrow several concepts from this work and indeed we build over a slightly modified version of their algorithm.
However, crowdsourced classification algorithms for us  are to a large extent swappable components we leverage: we do not aim at improving crowd-only classification but rather to study how to efficiently combine ML and human classifiers, and specifically how to do so in a manner that is robust to weak classifiers to avoid compromising on accuracy and avoid spending money in classifications not likely to produce good results.

\medskip 
 
Multi-filter classification has been also widely studied in search and information retrieval contexts. Indeed, an instance of such problem is that of selecting which predicates to apply first when filtering tuples in a query. The seminal work in this area is by Hellerstein and Stonebraker, discussing the problem of ranking predicates in query plans \cite{Hellerstein93predicates}. 

More recently, Franklin et al. \cite{CrowdDB2011}, Park and Widom \cite{ParkWidom13} and Parameswaran et al \cite{crowdscreen12,optimalWidom2014} focused on a similar  problem but in the context of crowd-powered databases, while Anvur et al \cite{Eddies2000}, Babu et al \cite{Babu2004adaptive}, Lan et al \cite{Lan_dynamicfilter_hcomp17} presented \textit{adaptive} extensions that tune the algorithm  as the query progresses (see Deshpande et al \cite{DeshpandeAdaptive} for a review).
We differ in many ways with respect to the prior art in this area. First, again, we aim at combining ML classifiers with crowd. 
Furthermore, we follow an  approach that makes initial estimates on the characteristics of the classification problem and then continuously revise the estimates, both in general and for each item to be classified, to identify an item and filter specific strategy to reduce errors and cost. Importantly, we identify the items on which we should give up trying, since it is more efficient to leave them for experts to screen.


Other prior work also addresses the issue of optimizations in terms of costs for obtaining labels and techniques to reduce cheating \cite{smyth1995inferring,karger2011budget,hirth2013analyzing,eickhoff2013increasing,hirth2011cost}. For example, Hirth and colleagues \cite{hirth2013analyzing} recommend specific cheating detection and task validation algorithms based on the cost structure of the task. 
In this paper we do not optimize cheating detection or even task design to a large extent, in that those are orthogonal aspects with respect to the methods and optimizations discussed here.

\subsection{Hybrid Classification}
Just like crowd classifiers, ML classifiers are also ingredients of a solution we borrow from the state of the art. In the context of our problem the interesting issues are how to combine ML and humans, how to ensemble available classifiers, and how to balance the request cost of (crowd or expert) votes with the need of training and testing ML models.

Hybrid classification is  an increasingly active area of research (see \cite{Vaughan_crowd17} for a recent survey).
Many researchers study the problem in the context of clustering or entity resolution (\cite{crowdER_2012,VesdapuntER_vldb2014,Vinayak_crowdcluster_nips2016,Gomes_crowdclustering_nips2011}).  
In general, many of these techniques operate by first classifying items automatically when ML classifiers are highly confident, and  by then leveraging ML to shape the kind of task proposed to the crowd (e.g., identifying potential clusters or matching items to propose to the crowd \cite{VesdapuntER_vldb2014,Vinayak_crowdcluster_nips2016}).
Another class of hybrid approaches, also extremely popular in industrial applications, are those where ML makes a proposal or a pre-filtering and humans confirm or refine. This happens in many fields, recently even in fashion \footnote{https://multithreaded.stitchfix.com/blog/2016/03/29/hcomp1/}. 

Kamar and colleagues propose instead a promising approach where crowd features (votes, as well as potentially other aspects of the crowdsourcing process such as task completion times) and task features are combined into a broader set of features to be used to learn a model~\cite{Kamar_2012_combining}. 
Researchers also explored using the crowd to extract features and patterns to then be leveraged by classifiers ~\cite{flock_2015,Carlos_pattern}.

While very interesting, these results are complementary to the work discussed here as we do not aim at developing a good base classifier but at leveraging effectively a set of arbitrarily weak classifiers we are given for the problem at hand. 
Perhaps the most closely related prior art is the work by Nguyen et al. \cite{Nguyen2015}, who adopt a clever mix of crowd+expert+machine learning  with an active learning classifier, where papers to be labeled are iteratively chosen to minimize overall cost and loss, by comparing estimated loss of crowd classification versus expert classification. Our problem and approach differs, however, in that we focus on exploiting existing classifiers and in doing so for optimally selecting the (item,filter) pair to poll the crowd on. 
It is indeed in the optimization of which question to ask the crowd that lies the essence of the approach we propose.